\documentclass[letterpaper]{article} 
\usepackage{aaai25}  
\usepackage{times}  
\usepackage{helvet}  
\usepackage{courier}  
\usepackage[hyphens]{url}  
\usepackage{graphicx} 
\usepackage{natbib}  
\usepackage{caption} 
\usepackage{algorithmic}
\usepackage[linesnumbered,ruled,vlined]{algorithm2e}
\usepackage{amsmath,amssymb,amsfonts}
\usepackage[nameinlink]{cleveref}
\usepackage{optidef}
\usepackage{booktabs}
\usepackage{colortbl}
\usepackage{xcolor}
\usepackage{paralist}
\usepackage{hhline}

\newtheorem{definition}{Definition}

\newtheorem{observation}{Observation}
\newtheorem{constraint}{Constraint}

\title{Influence Maximization in Temporal Social Networks with a Cold-Start Problem: A Supervised Approach}

\author {
    Laixin Xie\textsuperscript{\rm 1}\thanks{This work was completed during an internship at NetEase.},
    Ying Zhang\textsuperscript{\rm 1},
    Xiyuan Wang\textsuperscript{\rm 1}, Shiyi Liu\textsuperscript{\rm 2}, Shenghan Gao\textsuperscript{\rm 1},\\
    Xingxing Xing\textsuperscript{\rm 3}, Wei Wan\textsuperscript{\rm 3}, Haipeng Zhang\textsuperscript{\rm 1}, Quan Li\textsuperscript{\rm 1}\thanks{Corresponding author.}\\
}
\affiliations {
    \textsuperscript{\rm 1}School of Information Science and Technology, ShanghaiTech University\\
    \textsuperscript{\rm 2}Arizona State University\\
    \textsuperscript{\rm 3}UX Center, NetEase Game\\
    \{xielx, zhangying12022, gaoshh1,zhanghp,liquan\}@shanghaitech.com, \\
    shiyiliu@asu.edu,\\
    \{xingxingxing, gzwanwei\}@corp.netease.com
}

\usepackage{bibentry}

\begin{document}

\maketitle

\begin{abstract}
Influence Maximization (IM) in temporal graphs focuses on identifying influential ``seeds'' that are pivotal for maximizing network expansion. We advocate defining these seeds through Influence Propagation Paths (IPPs), which is essential for scaling up the network. Our focus lies in efficiently labeling IPPs and accurately predicting these seeds, while addressing the often-overlooked cold-start issue prevalent in temporal networks. Our strategy introduces a motif-based labeling method and a tensorized Temporal Graph Network (TGN) tailored for multi-relational temporal graphs, bolstering prediction accuracy and computational efficiency. Moreover, we augment cold-start nodes with new neighbors from historical data sharing similar IPPs. The recommendation system within an online team-based gaming environment presents subtle impact on the social network, forming multi-relational (i.e., weak and strong) temporal graphs for our empirical IM study. We conduct offline experiments to assess prediction accuracy and model training efficiency, complemented by online A/B testing to validate practical network growth and the effectiveness in addressing the cold-start issue.
\end{abstract}

%

\section{Introduction}
\par Virtual social networks like Twitter are pivotal extensions of physical social relationships, connecting billions of users and profoundly shaping human society. Influence Maximization (IM)~\cite{li2023survey} aims to expand network scales within a given diffusion model. IM finds diverse applications, from interrupting the spread of COVID-19~\cite{cheng2020outbreak} to enhancing viral marketing strategies~\cite{castiglione2020cognitive} and facilitating rumor control~\cite{vega2020influence}. At its core, IM identifies active members, or ``seeds'', with robust propagation capabilities, initiating the diffusion process within the network. Strategically pinpointing and leveraging these influential nodes enables maximal impact within the network, serving various societal and strategic objectives. 

\par When applied to temporal graphs~\cite{yanchenko2023influence}, IM maintains its primary objective while accounting for the chronological order of edge establishment, introducing complexity due to evolving connections over time. 
IM is known for its NP-hard nature, and the greedy algorithm is a traditional method for approximating the optimal solution~\cite{yanchenko2023influence}. In contrast, \textit{SPEX}~\cite{li2021spex} motivates us a supervised pipeline aimed at expanding the network scale, which is the ultimate goal of IM. This approach explicitly defines active members, thereby bypassing the need to address the NP-hard problem directly. To be specific, we adopt Influence Propagation Path (IPP) from SPEX, wherein the initial node forms social connections with multi-hop friends over consecutive timestamps. In the context of the IM problem, the initial node of any IPP inherently qualifies as an ``active member''. In this supervised context, accurately predicting these labeled active members becomes crucial for network expansion. Thus, the challenges of \textit{efficiently labeling IPPs and accurately predicting active members} ($\boldsymbol{\Gamma1}$) emerge as central considerations. In addition, the IM problem faces the often-overlooked cold-start issue, characterized by sparse social information at the outset, leading to performance declines for initial nodes. During the diffusion process, the initial network scale is smaller than the final scale, indicating the presence of cold-start nodes. This issue recurs across timestamps in temporal graphs, adds complexity. Essentially, \textit{the IM problem in temporal graphs is significantly affected by a severe and persistent cold-start problem} ($\boldsymbol{\Gamma2}$). Despite extensive exploration in social recommendation~\cite{panda2022approaches}, the cold-start issue has received limited attention in the context of IM.

\par In this study, we tackle the above two challenges ($\boldsymbol{\Gamma1}$ and $\boldsymbol{\Gamma2}$) with the following solutions. First, to address the first challenge ($\boldsymbol{\Gamma1}$), we introduce Motif-Based Filtering (MF), streamlining the identification of all IPPs by using known initiators as ground truth. Additionally, we implement a Unified Temporal Graph Network (TGN), specifically tailored for multi-relational temporal graphs. This TGN incorporates edge establishment, conversion, and diminishing, ensuring accurate predictions of active members. By transforming operations into equivalent tensor operations, our TGN implementation significantly enhances efficiency and scales effectively with larger datasets. Second, to address the cold-start problem ($\boldsymbol{\Gamma2}$), we efficiently augment cold-start nodes by adding new neighbors from historical timestamps that share similar IPPs. To expedite the matching of similar IPPs, we encode IPPs as strings and construct a prefix tree based on these strings.

\par In our experimental design, we conducted an empirical study through A/B testing within an online team-based game, utilizing our IM solution to enhance the game's native social recommendation system's coverage. The game's temporal multi-relational graphs, formed by edge timestamps and strong and weak relationships, perfectly align with our desired scenario. Augmenting the game's network scale emerges as a pivotal factor in boosting the impact of its built-in social recommendations. We selected this game due to its robust socialization features, which facilitated comprehensive data collection and empirical study. The offline experiment systematically highlighted the prediction accuracy of our approach, while the online experiment evaluated real-world network scale growth. Results from both experiments consistently affirm our method's superior performance. In the online experiment, our method demonstrates a \underline{$3.52\%$} overall improvement and a \underline{$14.32\%$} improvement in cold-start scenarios for spreading, compared to all other evaluated methods. In summary, our contributions are as follows: 
\begin{itemize}
    \item \textbf{Efficient Labeling Methods:} We focus on a subset of active members and introduce efficient labeling methods to enhance IM adaptability. Additionally, our implementation achieves up to a 22-fold speedup in TGN training under our experimental settings, providing a practical alternative to current high-performance GNN frameworks.
    \item \textbf{Addressing the Cold-Start Issue:} We propose a method to efficiently provide neighborhood information for cold-start nodes, effectively tackling this issue in IM.
    \item \textbf{Practical Application and Validation:} We apply our approach to a specific practical problem, demonstrating its effectiveness in expanding network scale through A/B testing in an online experiment.
\end{itemize}


\section{Related Work}
\subsection{Influence Maximization (IM) and Temporal Graph}
\par IM involves a diffusion model delineating the diffusion process and an algorithm tailored to identify active members. The diffusion simulation typically relies on models like the Independent Cascade (IC) or Linear Threshold (LT)~\cite{li2023survey}. These models estimate the probability of diffusion based on data from the downstream domain, simulating the network's post-diffusion state. Greedy algorithms are commonly employed to seek suboptimal solution for this NP-hard problem~\cite{feng2023influence,zhang2023capacity}. Our focus lies on addressing the IM problem within temporal graphs, particularly exploring supervised solutions. IM in temporal graphs accounts for temporal variations by considering both the formation and disappearance of edges. Consequently, it grapples with challenges posed by NP-hard complexities and graph dynamics. Recent surveys~\cite{yanchenko2023influence} include various traditional solutions, approximation algorithms~\cite{erkol2020influence}, and heuristic node ranking methods~\cite{michalski2020entropy}, which diverge from our proposed solutions.

\subsection{Deep Learning (DL) for IM}
\par DL solutions for IM have gained prominence due to their effectiveness in handling large-scale networks. In static graphs, Wang et al.~\cite{wang2021influence} conducted empirical studies employing graph embedding via unsupervised learning. They initially employed \textit{struc2vec}~\cite{Ribeiro2017struc2vecLN} to generate node embeddings with structural semantics, followed by clustering techniques to identify active members. For temporal graphs, recent approaches in dynamic IM (i.e., IM on temporal graph), as discussed in a survey~\cite{li2023survey}, draw inspiration from either the reinforcement learning (RL) framework~\cite{pmlr-v139-meirom21a,Mendona2020EfficientID} or from updating embeddings based on graph dynamics~\cite{Peng2021DynamicIM}. In supervised learning, methods either utilize proxies for active members~\cite{Qiu2018DeepInfSI} or generate labels based on diffusion models~\cite{kumar2022influence,kumar2023influence} before applying an improved prediction model. Our supervised learning approach builds on existing research for label generation but introduces a novel element by incorporating multi-relational temporal graphs.

\subsection{Research Gap}
\par This study addresses several research gaps within the domain of IM on temporal graphs. Notably, there has been limited exploration of supervised learning methods in this area. Our approach adheres to the traditional IM framework but introduces a novel labeling technique based on motif identification, effectively addressing our first set of challenges. Moreover, the cold-start issue has received little recent attention in the IM domain~\cite{panda2022approaches}. Existing solutions~\cite{ojagh2020social,herce2020new,bedi2020combining,wang2020cdlfm} often rely on supplementary information beyond the primary data. Temporal-aware approaches~\cite{wei2017collaborative,zhang2015social,chalyi2019method} primarily focus on user-item scenarios. In contrast, our solution innovatively acquires neighbors from observed timestamps, utilizing solely the inherent temporal social networks. Consequently, this cold-start solution for the IM problem addresses our second set of challenges.

\section{Preliminary and Background}
\par In real-world scenarios, social relationships among people change dynamically over time. At a given timestamp $t$, a social network is represented as $G_t=(V_t,E_t)$, where $V_t$ denotes the individuals (nodes) and $E_t$ denotes their relationships (edges). For any edge $e^t_{i,j}\in E^t$, it connects nodes $v^t_i$ and $v^t_j$ within $V_t$. If we consider one week as a time unit, a temporal graph includes all nodes and edges that occur within that week. The neighbors of a node $v^t_{i}$ at timestamp $t$ are defined as $N^t_i=\bigcup_{\forall e^t_{i,j}} v_j$, which includes all nodes connected to $v_i$ at timestamp $t$. Furthermore, we define $N^t(V)=\bigcup_{v_i \in V} N^t_i$ to facilitate subsequent discussions.

\par \textbf{Diffusion and Influence Maximization.} In a social network, diffusion refers to a recursive propagation process from a node to its neighbors, forming a rapidly expanding diffusion network. Over timestamps from $0$ to $T-1$, the network scale is defined as $\sigma(S, T-1)=|\bigcup_{t=0}^{T-1} V_t|$, where $S=V_0$. The goal of IM is to maximize the network scale by selecting seeds $S$ (i.e., active members) as the initial nodes of $G_0$. For simplicity, we start our timestamp from $0$, and the number of seeds remains constant to ensure a fair evaluation. Therefore, the IM problem on a temporal graph can be generally defined as: $S^*=\mathop{\arg\max}\limits_{|S|=K} \sigma(S,T)$, where $K$ is the fixed number of seeds~\cite{yanchenko2023influence}. 

\par \textbf{In-game Recommendation and Propagation.} On the gaming platform, we focus on team creation through the \textit{in-game teammate recommendation system} to study how its exclusive contribution to teaming behavior. This feature is fundamental across various game modes, supporting both one-on-one team fights and multi-team battles. The recommendation system suggests compatible teammates, thereby enhancing the gaming experience and improving player retention. 
Subsequently, three types of social relationships are derived: \textbf{exposure, invitation and adoption}. Exposure occurs when players appear on someone's recommendation list, allowing invitations to be sent. If an invitation is accepted, adoption occurs, and the invited player joins the inviter's team.

\section{Motivated Insights}
\par
\begin{figure}[t!]
    \centering
    \includegraphics[width=\linewidth]{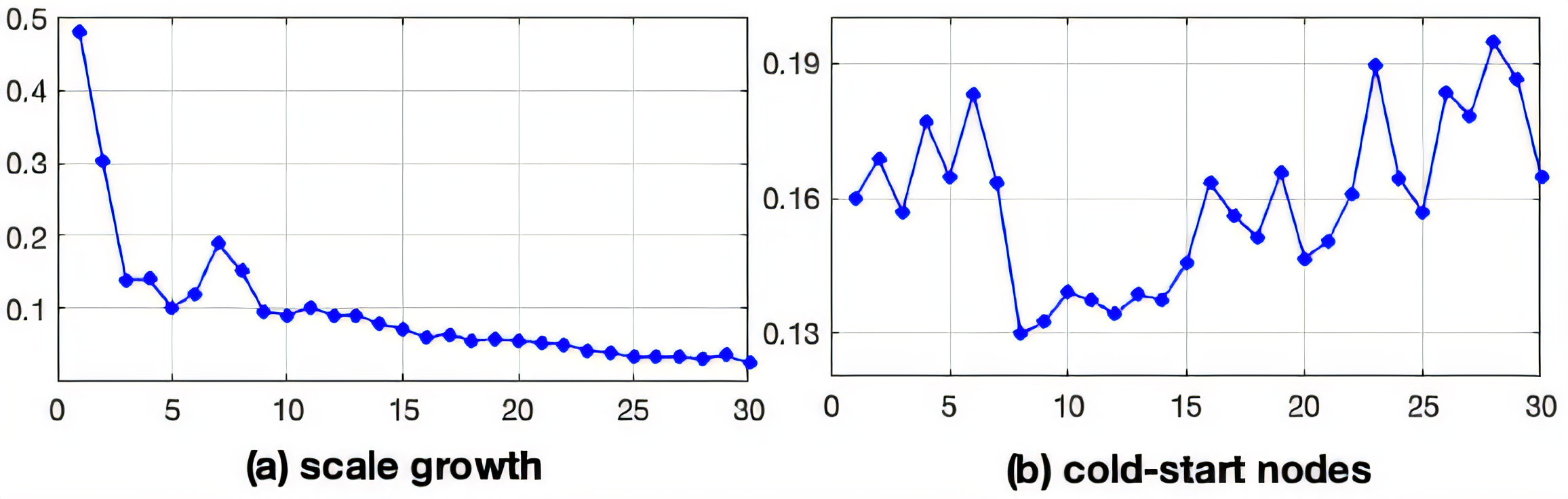}
    \caption{In the graph constructed from deduplicated edges derived from daily invitations and adoptions, the proportion of nodes with only one neighbor over the total number of nodes (y-axis) is presented over a span of $30$ days (x-axis).}
    \label{fig:obs1}
\end{figure}

 Data from a team-versus-team game developed by NetEase Games\footnote{NetEase Games is a leading global developer and publisher of video game IP across a variety of genres and platforms.} indicates that only 25.2\% of players used the social recommendation system (sending invitations or adopting system recommendations) over a month-long period. This low engagement suggests that conventional social recommendations are inadequate for most players. Additionally, network scale analysis during the same period shows a significant decline in growth rate: an initial $47.89\%$ increase drops to $12.05\%$ after the first week and further reduces to $2.5\%$ by the end of the month, as shown in Figure~\ref{fig:obs1}(a). This trend suggests the network scale will eventually stabilize and converge to a constant value. These findings imply that the recommendation system only effectively reaches a limited number of players, resulting in a minimal impact on the overall social network.
\begin{observation}
\textbf{LIMITED-COVERAGE RECOMMENDATION}: The recommendation system has a limited impact on the entire social network.
\end{observation}

\par Our investigation into the evolution of neighbor count across timestamps reveals a persistent ``cold-start'' issue in temporal graphs. As the network scale expands during 
diffusion in the IM problem, the number of neighbors for each node increases. However, the ratio of nodes with only one neighbor for social relationships ranges from $12.97\%$ to $19.44\%$, as depicted in Figure~\ref{fig:obs1}(b). This underscores that nodes with few neighbors, reflective of the cold-start issue, constitute a substantial portion of the network. Such nodes can hinder performance when methods depend on social information from their neighborhood. Contrary to the assumption that the cold-start issue is confined to initial timestamps, our findings demonstrate its persistence throughout propagation in IM problems, leading to compromised diffusion estimation for affected nodes.

\begin{observation}
\textbf{PERSISTENT COLD-START}: The consecutive presence of the cold-start issue across timestamps results in an underestimated propagation process.
\end{observation}

\section{Problem Definition}
\par The observed decline in growth, as highlighted in \textit{Observation 1}, is attributed to recommending acquaintances rather than players with the potential to expand the network. Such recommendations often lead to internal propagation, commonly known as the ``information cocoon''~\cite{peng2021breaking}. Consequently, \textit{Observation 1} prompts us to focus on recommending specific types of active players who can facilitate broader propagation. This supervised approach differs from most existing IM solutions, which often fail to explain the patterns crucial for expanding network scale. To introduce external propagation and break the cocoon, we classify relationships formed through recommendation-driven invitations/adoption as ``strong'' relationships, while the exposure in the system is deemed ``weak'' (\textit{Definition \ref{def:weak&strong}}). Any strong relationships must occur after exposure, making the weak graph an upper bound (\textit{Constraint \ref{constraint1}}) for IM solutions.

\begin{figure*}[h]
 \centering 
 \includegraphics[width=\textwidth]{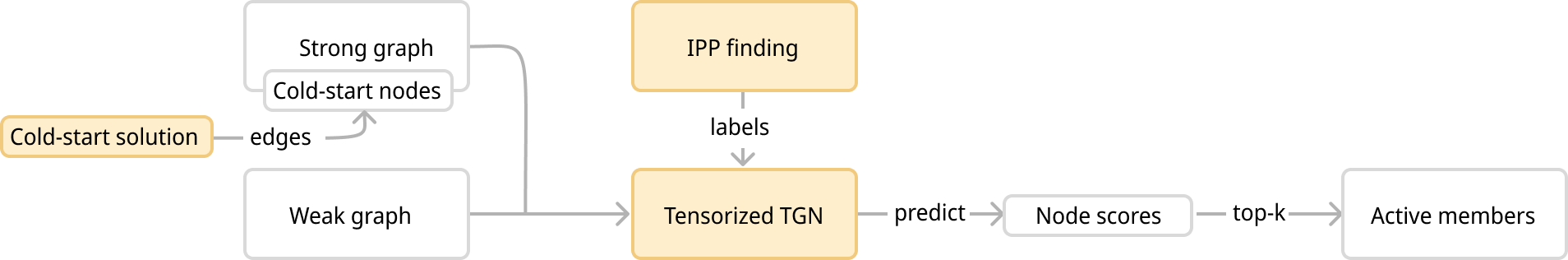}
 \caption{Pipeline for predicting active members. Weak and strong graphs are constructed using exposure edges and invitation and adoption edges derived from in-game teammate recommendations, respectively. These graphs are then input into a Tensorized TGN to generate node scores. During training, node scores with labels from IPP findings are used to calculate the training loss. In the inference phase, nodes with top scores are active members.}
 \label{fig:pipeline}
\end{figure*}

\begin{definition}
\label{def:weak&strong}
Given a temporal graph $G_t(V_t,E_t)$, the strong graph $G_t^s(V^s_t, E_t^s)$ is a subgraph where $V^s_t \subseteq V_t$ and $E_t^s \subseteq E_t$, such that $\forall e^t(v_i^t, v_j^t)\in E^s_t$, both $v_i^t$ and $v_j^t$ belong to $V_t^s$. Let $G_t^w(V_t^w,E_t^w)=G_t(V_t,E_t)$ represent the weak graph.
\end{definition}

\begin{constraint}
\label{constraint1}
\label{ctt:upper-bound}
$\left|\sigma\left(S^*, T\right)\right|\leq\left|\bigcup_{t=0}^{T-1} V_t^w\right|,S^* \in V_0^s$, where $S^*$ is the IM solution.
\end{constraint}

\par The distinction between strong and weak relationships has been extensively studied in previous literature~\cite{granovetter1973strength,onnela2007structure}, denoting high and low establishment probabilities. Traditional recommendation systems typically consider strong relationships as positive edges and weak relationships as negative edges in unsupervised learning~\cite{hamilton2017inductive}. However, it's important to recognize that strong relationships can evolve from weak ones through players' invitation/adoption behavior. Weak relationships play a crucial role in setting an upper bound for the network scale achievable by IM solutions.

\par Inspired by \textit{Observation 2}, we formulate the cold-start issue as another constraint for the IM problem, shaping our approach to addressing this challenge.

\begin{definition}
Given a strong graph $G_t^s(V^s_t, E_t^s)$, we define a cold-start node as $\exists c^t\in V_t^s$ such that $|N^t(c_t)|\leq C$, where $C$ is a constant (e.g., 1). Let $\widehat{C}_t$ represent the universal set that includes all $c^t$ at timestamp $t$.
\end{definition}

\begin{constraint}
\label{ctt:cold-start}
$\forall t \in[0, T-1],\left|\widehat{C}_t \right| \geqslant \delta$, where $\delta$ is the  minimal number of cold-start nodes through statistics over $T$ temporal graph.
\end{constraint}

These nodes are considered minimal because increasing $C$ would naturally include more nodes, thus broadening the definition of cold-start nodes.

\par Under these constraints, our IM problem is defined as:
\begin{maxi}
  {|S|=k}{\sigma(S, T)}{}{}
\addConstraint{\text{\textit{Constraint 1}}}{}
\addConstraint{\text{\textit{Constraint 2}}}{}
\end{maxi}
\par We argue that traditional metrics, such as adoption rates~\cite{zhang2023capacity}, do not effectively reach players outside the coverage of the recommendation system. Therefore, a primary focus of our research is to expand the limited coverage of the teammate recommendation system. These two objectives are not in direct competition.

\section{Algorithmic Solution \& Implementation}
\par In this study, we formulate the IM problem as a supervised learning task, shown in Figure~\ref{fig:pipeline}. In the following sections, we first introduce the problem formulation and the efficient labeling of IPP findings. Next, we provide an overview of the TGN architecture used for label prediction, highlighting modifications to accelerate training and adapt to large-scale datasets. Lastly, we explore the serialization of IPPs and an efficient retrieval method designed to address the cold-start issue.

\subsection{Supervised IM and Efficient Labeling}
\par Before formulating IM as a supervised problem, we first introduce IPPs within both the strong and weak graphs:

\begin{definition}
\label{def:ipp}
An Influence Propagation Path (IPP) is defined as a sequence of edges $<e_{v_0,v_1}^{t_0},e_{v_1,v_2}^{t_1},\dots,e_{v_{N_{p}-1},v_{N_{p}}}^{t_{N_{p}-1}}>$ that adheres to the edge condition $t_0\leq t_1\leq \dots \leq t_{N_{p}-1}$ and the node condition $\forall i\in [0, N_{p}-2], v_i\in V_{t_i}^s$, while $v_{i+1},v_{i+2}\notin V_{t_i}^s$. Here, $N_p$ signifies the length of the IPP.
\end{definition}

\par \textbf{Supervised IM}. We highlight an important modification from the original definition, which revolves around the node constraint regarding the strong and weak graph, derived from CONSTRAINT \ref{ctt:upper-bound}. This study particularly focuses on the simplest case where $N_p=2$, aiming to formulate a supervised learning problem using these IPPs as ground truth. We first train our model to predict IPPs, without seeking the IM solution for the observed data. Subsequently, during inference, the inference data are treated as the initial network. The trained model then predicts the seeds for maximizing the subsequent diffusion network.

\par \textbf{IPP Findings}. Identifying all IPPs consumes a considerable amount of time, prompting the use of a sampling strategy as a compromise to access a subset of labels, as demonstrated in a prior study~\cite{li2021spex}. To efficiently identify entire labels, we introduce a motif-based filtering (MF) method according to DEFINITION \ref{def:ipp}. The MF method begins with a motif finding algorithm aimed at detecting all $2$-hop paths $<e_{v_0,v_1}^{t_0},e_{v_1,v_2}^{t_1}>$. Subsequently, these paths undergo filtering based on two conditions: the edge condition ($t_0\leq t_1$) and the node condition ($v_0\in V^s_{t_0}$, while $v_1,v_2\notin V^s_{t_0}$). 
To efficiently determine a node's inclusion within $V_{t_0}^s$, a strategy is applied where each node initially receives a unique ID in ascending sequential order. The decision on a node's inclusion is based on whether its ID exceeds the maximum ID within $V^s_{t_0}$.

\begin{figure*}[h]
 \centering 
 \includegraphics[width=\textwidth]{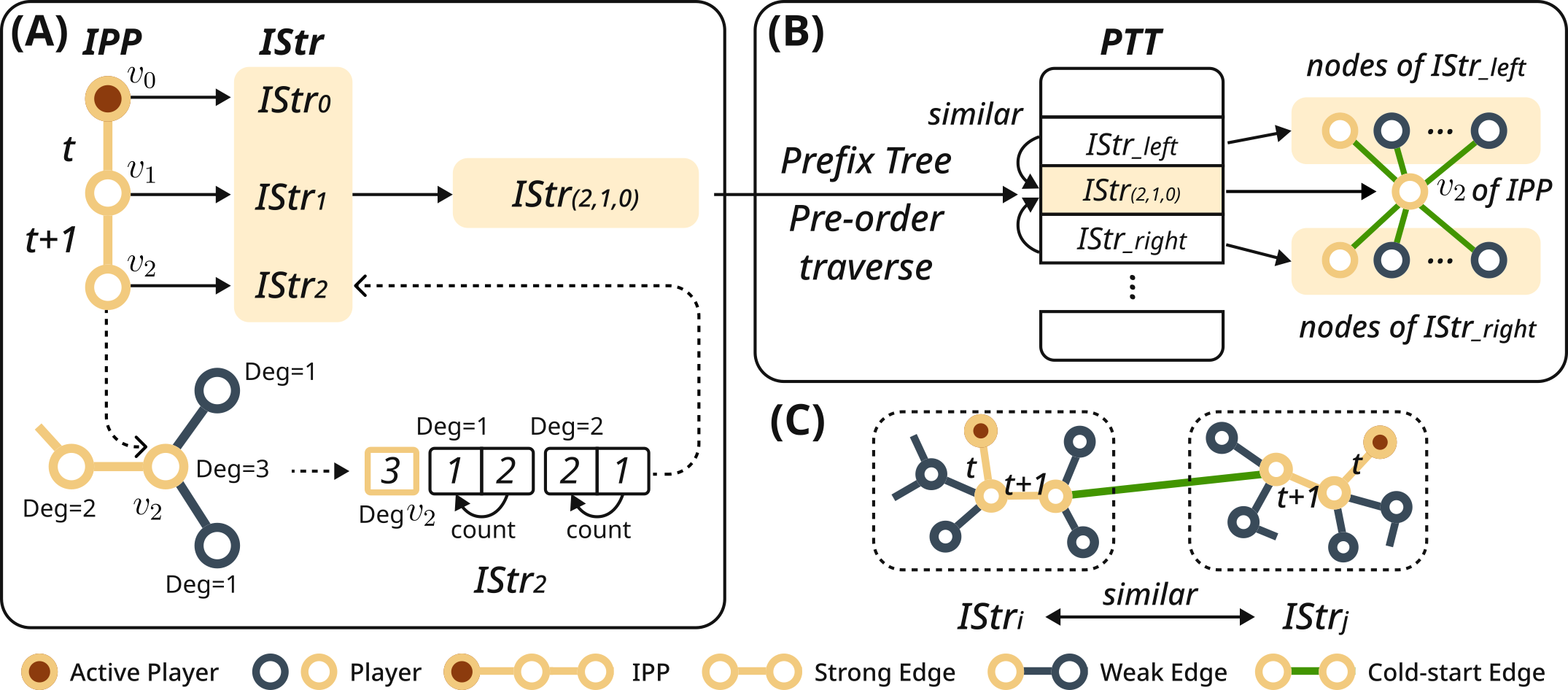}
 \caption{Cold-start solution: (A) IPP Serialization: Serialized strings are generated for each node in an IPP and then concatenate into an \textit{IStr}. (B) Neighbor Retrieval: The \textit{IStr}s of all IPPs are inserted into a prefix tree and positioned by pre-order traversal (PTT), ensuring that similar strings are placed adjacently. Cold-start edges are established between the end node $v_2$ of an IPP and other nodes belonging to the most similar \textit{IStr}s in the PTT. (C) Demonstration of neighbor retrieval.}
 \label{fig:coldstart_alg}
\end{figure*}

\par \textbf{Time Complexity.} The time complexity of our algorithm hinges on the motif finding algorithm employed. We argue that any motif finding algorithm's time complexity must be at least linear to $\mathcal{O}(\sum_{t=1}^T|Motif_t|)$, where $Motif_t$ represents identified motifs at timestamp $t$. This assertion holds because validating each edge's eligibility as the desired motif takes at least $\mathcal{O}(1)$, assuming these edges are precisely the final motifs in the best-case scenario. Both edge and node filtering processes involve iterating through all motifs, resulting in a time complexity of $\mathcal{O}(\sum_{t=1}^T|Motif_t|)$. Thus, the time complexity of the filtering process for IPPs is linear to the number of found motifs and must be less than the motif identification process. Our implementation employs \textit{Dotmotif}~\cite{Matelsky_Motifs_2021}, an efficient variant of the VF2 algorithm that decouples time complexity from the number of edges. Finally, compared with implementations that recursively examine nodes' neighbors and validate the formation of 2-hop paths, our motif-filtering method excels in handing dense graphs, where the edge count grows exponentially with the node count.

\subsection{TGN Deployment and Tensorization}
\par To predict IPPs, we implement TGN to fit in our scenario. Our implementation also tensorizes two modules (memory and aggregator) of TGN to improve efficiency.

\par \textbf{TGN Architecture}. TGN incorporates a specialized storage mechanism termed \textit{memory}, wherein a vector assigned to each node encapsulates the most recent data related to the edges. The training procedure of TGN is demonstrated through a batch comprising $e_{v_0,v_1}^{t_0},e_{v_1,v_2}^{t_1}$. Initially, node embeddings for $v_0^{t_0},v_1^{t_0},v_1^{t_1},v_2^{t_1}$ are generated by the embedding module, utilizing the temporal graph and the node's \textit{memory}. Subsequently, these embeddings, along with edge information, serve as historical \textit{messages} to update the memory of the nodes. Particularly, information for $v_1$ at $t_0$ and $t_1$ is aggregated, updating its \textit{memory} to $t_1$. Following this batch, the \textit{memory} stores three vectors for $v_0^{t_0}, v_1^{t_1}, v_2^{t_1}$. For more details, please refer to the paper~\cite{rossi2020temporal}.

\par \textbf{Specification}. In our scenario, the edge attributes are categorical, with values of $0$, $1$, $2$ representing weak relationships, strong relationships, and the additional connections (introduced in Section~\ref{subsec:cold-start_solution}) for cold-start nodes, respectively. This notation seamlessly aligns with the requirements of our temporal multi-relational network implemented by TGN. Subsequently, the generated node embeddings are passed through a Multi-layer Perceptron (MLP) decoder, consisting of two Linear-ReLU-Dropout blocks followed by a Linear layer, with an output dimension of 2. The decoder is then trained using a supervised Binary Cross-Entropy (BCE) loss, with the labels identified by IPPs fed into it. Finally, the weighted sum of the supervised BCE loss and the unsupervised TGN loss forms the multi-task loss. During inference, the class with the higher value is considered as the prediction.

\par \textbf{Tensorization}. In our deployment of TGN, we have identified a key challenge that hampers efficiency. Specifically, concerning message aggregation, TGN explores two alternatives: exclusively utilizing the latest message (LM) and calculating the mean of all messages (MM). An ablation study~\cite{rossi2020temporal} indicates that while LM slightly outperforms MM in accuracy, it incurs a threefold computational time. Nevertheless, the original implementation is designed for supporting various aggregators. Recognizing this as a potential avenue for efficiency enhancement, we exclusively employ LM as the aggregator and implement a tensorized \textit{memory} and LM aggregator to minimize redundant loops. This tailored implementation results in a significant improvement in efficiency, surpassing the aforementioned threefold acceleration. Moreover, our acceleration on a single operation is orthogonal to the parallel framework designed to accelerate general Temporal Graph Neural Networks (TGNNs)~\cite{xia2024redundancy,sheng2024mspipe,zhou2022tgl}, which is the primary focus of previous works.

\subsection{Cold-start Solution}
\label{subsec:cold-start_solution}

\par To address the neighbor insufficiency in cold-start nodes, we propose establishing connections among cold-start nodes that share similar IPPs. As shown in Figure~\ref{fig:coldstart_alg}, this process begins with the serialization of IPPs into fixed-length strings, facilitating subsequent calculation. After serialization, we construct a prefix tree for efficient retrieval of IPPs with similarities. This method adeptly manipulates social relationships within the network data, introducing edges that minimally alter the network's structure. Consequently, our solution to the cold-start problem aligns well with various algorithms designed for temporal graphs. The complete process is outlined in Figure~\ref{fig:coldstart_alg}.

\par \textbf{IPP Serialization}. We introduce degree counts for a specific degree $k$, denoted as:
\begin{equation}
    CD(v_i,k,t)=\sum_{nbr_j\in N^t{v_i}} \mathbb{I}(Deg(nbr_j, t)==k).
\end{equation}
Here, $Deg(v_i,t)$ denotes the degree of $v_i^t$, and $\mathbb{I}$ yields 1 when the condition is met. This serialization (Figure~\ref{fig:coldstart_alg}(A)) method preserves neighborhood information by associating each neighbor's degree with the degree count. To simplify, the serialization retains the node's degree and the three most frequent degree counts. To ensure consistent string length, degrees are truncated to $99$, with zeros appended for nodes lacking sufficient neighbors. Subsequently, for three nodes of an IPP $v_0^{t_0}, v_1^{t_1},v_2^{t_2}$, we generate serialized strings for each involved node, which are then concatenated in reverse order to form $IStr_{v_2,v_1,v_0}^{t_1,t_1,t_0}$. This reverse concatenation is specifically designed for the inference phase, where only historical data are available. The retrieval of IPPs thus focuses on identifying potential neighbors for the latest node $v_2$ of an IPP, utilizing the serialized IPPs.

\par \textbf{Neighbor Retrieval}. The Trie, or prefix tree, plays a crucial role in efficiently retrieving IPP strings based on common prefixes. After inserting all IPP strings into the Trie, strings with similarities are positioned adjacently within the sequence generated through the Trie's pre-order traversal, denoted as PTT (construction and traversal details omitted). Subsequently, for a node $v_i$, we illustrate the retrieval process of an IPP ending with $v_i$, and its serialization is simplified to $IStr_i$ (illustrated in Figure~\ref{fig:coldstart_alg}(B) and Algorithm~\ref{alg:retrieval}). Similar strings are placed adjacently to the position, where $PTT[position]=IStr_i$. These identified strings reflect potential neighbors for the given nodes. A subset of these nodes is randomly selected to augment $v_i$'s neighborhood. Additionally, we filter the identified strings based on similarity to the retrieval string. Specifically, considering each two words as a number within an IPP string ($N_{I}$ numbers in total), the similarity between any two IPP strings, $IStr_i$ and $IStr_j$, is calculated as:
\begin{equation}
    Sim(IStr_i, IStr_j)=\sum_{k=0}^{N_{I}-1} \mathbb{I}(IStr_i[k]==IStr_j[k]),
\end{equation}

\begin{algorithm}[ht]
\DontPrintSemicolon
\caption{Neighbor Retrieval (IPPs, PTT, w, h)}
\label{alg:retrieval}
  $E_c\leftarrow \emptyset;$\\
  \For{$(v_2,IStr)\in IPPs$}
  {
    $pos \leftarrow \text{Index}(PTT, IStr); Q \leftarrow \emptyset;$\\
    \For{$IStr^{\prime}\in PTT[pos-w, pow+w]$}
    {
        \If{Sim($IStr, IStr^{\prime})<h$}
        {
            $V^{\prime}\leftarrow StringToNode(IStr^{\prime});$\\
            $Q \leftarrow Q \bigcup V^{\prime};$
        }
    }
    $Q \leftarrow Sample(Q);$\\
    $E_c \leftarrow \{v_2\}\times Q;$
  }
  \Return{$E_c;$}
\end{algorithm}

where $IStr[k]$ denotes the $k$-th number. To boost retrieval efficiency, our implementation utilizes caching, capitalizing on the relatively small size of the PTT compared to the total number of IPPs. An example is demonstrated in Figure~\ref{fig:coldstart_alg}(C), omitting details.

\par \textbf{Time Complexity}. If $d_{avg}$ represents the average node degree, generating an $IStr_{v_i}^{t_i}$ requires $\mathcal{O}(d_{avg})$ operations to access the degree of each neighbor. Assuming $M$ IPPs found, the total time complexity for IPP serialization becomes $\mathcal{O}(Md_{avg})$. Neighbor retrieval involves Trie construction, PTT generation, and neighbor retrieval for each cold-start node. Trie construction takes $\mathcal{O}(LM)$, where $L$ denotes the length of serialized string. PTT generation, while worst-case $\mathcal{O}(LM)$, is significantly lower in practical scenarios. The neighbor retrieval exhibits linear time complexity in indexing all strings, assuming a constant number of adjacent positions are considered similar. This indexing requires the length of PTT to populate the cache, with a worst-case complexity of $\mathcal{O}(M)$ where each serialized IPPs is distinct. Similarly, retrieving nodes based on $IStr$ also has a worst-case time complexity of $\mathcal{O}(M)$ to assign all nodes to each $IStr$. In summary, our approach to addressing the cold-start issue has a time complexity of $\mathcal{O}(Md_{avg}+LM)$. In practice, we suggest $\mathcal{O}(Md_{avg})$ due to shared prefixes, reducing the effective length of Trie traversal.

\section{Offline Experiment}
\par The aim of the offline experiment is to validate the predictive accuracy of our approach. We utilize four predictive models detailed in \textit{SPEX}~\cite{li2021spex} and five datasets for comparison. Furthermore, we evaluate the efficiency of model training across various scales and computational platforms, conducting experiments in two distinct environments to account for scale variations. The code is opensource~\footnote{\url{https://github.com/laixinn/ICWSM25-Influence-Maximization/}}.

\begin{table}[h]
\centering
\begin{tabular}{@{}lccccc@{}}
\toprule
Name & \textit{Twit.} & \textit{NetE.} & \textit{B.C.} & \textit{W.T.} & \textit{S.O.} \\ 
\midrule
nodes          & 50.4K & 1.4M  & 274 & 85.9K  & 0.6M \\
edges          & 59.0K & 13.4M & 809 & 524.6K & 7.5M \\
weeks          & 9 & 8 & 266 & 223 & 86 \\
strong nodes   & 39.7K & 0.2M & 218 & 66.8K  & 0.5M \\
strong edges   & 37.4K & 0.5M & 201 & 292.0K & 2.8M \\
ground truth   & 8.8K & 47.9K & 77  & 9.8K   & 0.1M \\
\bottomrule
\end{tabular}
\caption{Overall Statistics ($K=10^3,M=10^6$).}
\label{tab:dataset}
\end{table}

\begin{table*}
\setlength\tabcolsep{3pt}
\centering
\begin{tabular}{llllllllllllllll} 
\hline
\multicolumn{1}{l}{}              & \multicolumn{5}{c}{AUC (\%)}                                                                                          & \multicolumn{5}{c}{ACC (\%)}                                                                                          & \multicolumn{5}{c}{AP (\%)}                                                                                            \\ 
\cmidrule(lr){2-6} \cmidrule(lr){7-11} \cmidrule(lr){12-16}
\multicolumn{1}{l}{}              & Twit.          & NetE.        & B.C.         & W.T.         & \multicolumn{1}{l}{S.O.}    & Twit.          & NetE.        & B.C.         & W.T.         & \multicolumn{1}{l}{S.O.}    & Twit.          & NetE.        & B.C.         & W.T.         & \multicolumn{1}{l}{S.O.}     \\ 
\hline
\multicolumn{1}{l}{DiffNet++}    & 64.23            & OOM              & 30.47          & 52.16          & \multicolumn{1}{l}{59.84}          & 59.62          & OOM              & 61.49          & 50.51          & \multicolumn{1}{l}{58.00}          & 72.81          & OOM              & 31.07          & 34.06          & \multicolumn{1}{l}{52.23}           \\
\multicolumn{1}{l}{DiffNet++\&spex} & 60.68          & OOM              & 57.35          & 52.16          & \multicolumn{1}{l}{74.79}          & 53.56          & OOM              & 61.49          & 80.43          & \multicolumn{1}{l}{57.90}          & 64.16          & OOM              & 48.21          & 28.58          & \multicolumn{1}{l}{65.49}           \\ 
\multicolumn{1}{l}{FuseRec}       & 66.09          & OOM              & \textbf{77.46} & 57.50          & \multicolumn{1}{l}{63.40}          & 63.26          & OOM              & \textbf{77.84} & 45.57          & \multicolumn{1}{l}{55.37}          & 69.62          & OOM              & \textbf{80.28} & 68.43          & \multicolumn{1}{l}{76.61}           \\
\multicolumn{1}{l}{FuseRec\&spex}   & 66.30          & OOM              & \underline{77.38}  & 57.46          & \multicolumn{1}{l}{63.46}          & 62.02          & OOM              & \underline{77.84}  & 45.57          & \multicolumn{1}{l}{54.94}          & 69.59          & OOM              & \underline{80.12}  & 68.38          & \multicolumn{1}{l}{76.87}           \\ 
\multicolumn{1}{l}{GraphMAE}      & 76.93          & 81.72          & 61.35          & 53.00          & \multicolumn{1}{l}{50.42}          & 53.52          & \textbf{79.78} & 71.88          & 72.16          & \multicolumn{1}{l}{75.67}          & 37.73          & 76.02          & 46.48          & 29.94          & \multicolumn{1}{l}{24.49}           \\
\multicolumn{1}{l}{GraphMAE\&cold}  & 74.77          & \underline{83.16}  & 74.44          & 57.46          & \multicolumn{1}{l}{50.38}          & \textbf{87.25} & \underline{78.33}  & 75.00          & 72.32          & \multicolumn{1}{l}{75.68}          & 34.39          & 75.90          & 55.41          & 39.40          & \multicolumn{1}{l}{24.26}           \\ 
\multicolumn{1}{l}{TGN}          & \underline{77.67}  & 81.66          & 74.53          & \underline{96.33}  & \multicolumn{1}{l}{\underline{91.71}}  & 71.38          & 76.00          & 51.70          & \underline{91.61}  & \multicolumn{1}{l}{\underline{85.13}}  & \textbf{80.72} & \underline{80.55}  & 75.96          & \underline{97.58}  & \multicolumn{1}{l}{\underline{95.38}}   \\
\multicolumn{1}{l}{TGN\&cold}       & \textbf{78.24} & \textbf{83.48} & 74.81          & \textbf{97.30} & \multicolumn{1}{l}{\textbf{93.62}} & \underline{73.85}  & 77.20          & 51.70          & \textbf{92.99} & \multicolumn{1}{l}{\textbf{87.19}} & \underline{80.50}  & \textbf{83.72} & 77.05          & \textbf{98.26} & \multicolumn{1}{l}{\textbf{96.42}}  \\ 
\hline
\end{tabular}
\caption{IPP Prediction Performance (OOM: out-of-memory).}
\label{tab:offline_pred}
\end{table*}

\subsection{Experimental Setting}

\par \textbf{Dataset}. We utilize two datasets obtained from different platforms. A temporal graph is generated by aggregating interactions over a one-week timeframe, where the ground truth of IPP and addition edges for cold-start nodes are identified. The identified IPP is based on a 2-hop structure under the most relaxed conditions. This is because IPPs with longer hops must be a subset of the 2-hop IPPs, making these 2-hop IPPs representative. Detailed statistics are provided in Table~\ref{tab:dataset}, which aggregates data over time.

\par \textit{NetEase (NetE.)}. This dataset contains interaction data, considered as strong and weak relations, and player profiles obtained from an in-game recommendation system, spanning from January 3 to February 28, 2022. Experiments with this dataset are conducted on a high-performance machine featuring an AMD EPYC 7543 32-Core CPU, 377GB RAM, and an NVIDIA 24G A30 GPU.
\par \textit{Twitter (Twit.)}~\cite{lou2013learning,li2021spex}. This dataset comprises retweet logs from January 1 to March 1, 2010. Strong and weak relational edges are established based on a predefined threshold (0.3 quantile) on edges' accumulated weights. We utilize a BERT-base model\footnote{https://github.com/google-research/bert} to generate tweet embeddings, which are then aggregated to derive user features. Experiments on this dataset are conducted on a system equipped with an Intel(R) Xeon(R) Gold 5218 CPU, 512GB RAM, and an NVIDIA 24G GeForce RTX 3090 GPU.
\par \textit{SNAP Temporal Networks}\footnote{\url{https://snap.stanford.edu/data/index.html#temporal}}: BitCoins (B.C.), WikiTalk (W.T.), StackOverflow (S.O.)~\cite{kumar2018bitcoin,paranjape2017stackoverflow}. We leverage publicly available datasets from SNAP, encompassing a diverse range of temporal networks over extended periods. To better capture long-term propagation and ensure the resulted IPP labels, we aggregate each network into 60-day intervals. When no inherent node features are provided, we substitute them with random embeddings.

\par \textbf{Baseline}. In our evaluation, we consider the following baseline algorithms from two perspectives:

\par \textit{SPEX Competitors}. \textit{FuseRec}~\cite{narang2021fuserec}, \textit{DiffNet++}~\cite{wu2020diffnet++}, and their \textit{SPEX} adaptations, without scalable implementation, serve as prediction baselines in Twitter dataset. We adaptively replace items by users and modify supervised loss similar to \textit{SPEX}. The accuracy threshold is set at $0.5$. This baseline does not scale effectively to large datasets. Therefore, we intentionally downsample the \textit{S.O.} dataset to evaluate its maximum capable performance, by $50\%$ for FuseRec (\&spex) and $40\%$ for DiffNet++ (\&spex).

\par \textit{Graph Solutions}. \textit{GraphMAE} and \textit{TGN} are representative for the algorithms in static and temporal graphs. Their scalable implementations are utilized in both Twitter and NetEase. dataset. Both approaches employ a 2-class prediction.

\par For consistency, methods integrating our cold-start solutions or the \textit{SPEX} approach are labeled as ``\&cold'' and ``\&spex'', respectively. To maintain data integrity, datasets are split in a 4:1:1 ratio for training, validation, and testing, preserving temporal sequence integrity. Evaluation metrics including Average Precision (AP), ROC, AUC, and accuracy are aligned with TGN's setting.

\subsection{Result Analysis}
\par The experimental results, summarized in Table~\ref{tab:offline_pred}, emphasize the highest (\textbf{first}) and second highest (\underline{second}) scores. In most scenarios, \textit{TGN\&cold} achieve the best or near-best results, outperforming the \textit{spex} plugin designed for IPP identification. Several significant insights can be found from the results:

\par \textbf{\textit{Temporal Information's Impact}: temporal data inclusion markedly improves model performance.} In W.T. and S.O., two \textit{TGN}s consistently outperform in three metrics, averaging $95.46\%,90.09\%,97.34\%$. Conversely, moving from the short-term (Twit., NetE.) to long-term datasets (W.T., S.O.) reveals an enhanced performance for \textit{TGN\&cold} but a marked decline for two \textit{GraphMAE}s, particularly in AUC ($79.15\%$ to $52.82\%$ on average). This contrast underscores the critical role of temporal structure in delivering accurate IPPs for the following empirical study.

\par \textbf{\textit{Propagation-insufficient Scenario}: sparse propagation results in insufficient labels and degrades the performance for our cold-start solution.} In the B.C. dataset, $43.5\%$ timestamps include $\leq 1$ propagation (i.e., label). Decreasing the accumulation period further increases this proportion and exacerbates label imbalance. These observations illustrate the nature of the rare propagation behavior in the bitcoin scenario. Consequently, \textit{TGN}'s ACC drops to a secondary level, which carries over to the \textit{TGN\&cold}. In contrast, W.T. and S.O datasets have $100\%$ and $84.6\%$ of timestamps including more than $10\%$ of propagation, with average propagation proportions of $14.1\%$ and $22.0\%$. In these datasets, two \textit{TGN}s outperform other baselines in these datasets, demonstrating the strong correlation between propagation and performance. However, in the propagation-insufficient B.C. dataset, our cold-start solution is particularly helpful for \textit{GraphMAE}, with an average improvement of $8.38\%$. As discussed earlier, the static \textit{GraphMAE} struggles with learning temporal information, and thus the absence of propagation has a neutral or even positive effect on its performance. Overall, \textit{TGN\&cold}’s worst performance in B.C. arises from \textit{TGN}’s limitation in propagation-insufficient scenarios, while the cold-start solution itself remains benefit for static GNNs.

\par \textbf{\textit{Plugin-perspective Performance}: our cold-start solution (\textit{cold}) outperforms the plugin competitor SPEX (\textit{spex}) regarding overall stability.} These plugins share advantages of requiring no architectural changes to the base models. Integrating \textit{spex} and \textit{cold} delivers a state-of-the-art overall performance improvement of $3.20\%$ and $3.25\%$. However, \textit{cold} demonstrates a positive effect in $73.3\%$ of cases ($22$ out of the $30$), outperforming \textit{spex}'s $33.33\%$ ($8$ out of $24$ due to OOM). Notably, these results consider the integration to both static (\textit{GraphMAE)} and dynamic (\textit{TGN}) GNN, demonstrating the cold-start solution's broad applicability. Despite \textit{GraphMAE} showing a lower compatibility with \textit{cold} ($60\%$ positive effect) compared to \textit{TGN} ($93.3\%$), \textit{cold} still surpasses \textit{spex} when integrating to \textit{GraphMAE}. The observed variation in compatibility is attributed to the absence of temporal information, which affects model's capability in learning cold-start edges.

\subsection{Computational Efficient}
\par We conduct a comprehensive benchmarking on TGB~\cite{huang2023temporal} to demonstrate our efficiency improvement, as shown in Table~\ref{tab:efficiency}. The results indicate a speedup of up to $4\times$ faster than the TGN implementation. The first five datasets are for link prediction, and the subsequent four datasets are for node prediction. The experiment code is available~\footnote{\url{https://github.com/laixinn/TGB}}. Additionally, in the Twit. dataset, an ablation study examining the effect of batch size on efficiency enhancement showed that the per-epoch time decreases from $295$s to $112$s and from $162$s to $51$s for batch size of  $256$ and $1024$, respectively. This acceleration demonstrates a performance gain increasing from the raise of $1.82$ to $2.2$ when increasing the batch size fourfold, highlighting our method’s potential to scale even further. In the NetE. dataset, due to GPU memory constraints, the most efficient tensorized version of \textit{TGN} ($1.54$ hours for a batch size of 2048) is $22\times$ faster than its original counterpart ($34.27$ hours for a batch size of 1024). These observations collectively suggest that our accelerated approach yields potential performance benefits in large-scale scenarios. However, the acceleration provided by our implementation primarily serves as an orthogonal alternative and does not match the hundred-fold speedups offered by specialized parallel architectures~\cite{yu2024rtga}.

\section{Empirical Study}
\par We perform an A/B testing on NetEase Game's team-vs-team game platform to demonstrate the effectiveness of our method in addressing the IM problem. This online experiment ran from January 29 to February 3, 2024. At the beginning of each week, the model was updated with the latest monthly data.

\begin{table*}[h!]
\centering
\begin{tabular}{llllllllllllll}
\toprule
Implementation & Coin   & Wiki & Review & Comment & Flight & Trade & Token  & Genre & Reddit \\
\midrule
Original  & 472          & 2    & 198         & 1523         & 472          & 2          & 438          & 71          & 345          \\
Tensorized  & \textbf{204} & 2    & \textbf{35} & \textbf{326} & \textbf{470} & \textbf{1} & \textbf{260} & \textbf{50} & \textbf{207}    \\
\bottomrule
\end{tabular}
\caption{Time consumption (/s per epoch) on TGB.}
\label{tab:efficiency}
\end{table*}

\subsection{A/B Testing Environment}
\par We start by outlining the fundamental rules of the pre-defined in-game recommendation system on our platform. Our aim is to target invitations exclusively to online players, requiring daily training of our recommendation system's algorithms to suggest from a pool of real-time online players. We apply additional filtering rules, such as matching preferred game mode with the inviter, to ensure the relevance of recommendations. Subsequently, a portion of online players receive experimental recommendations. We highlight some key settings to ensure test fairness:
\par \textit{Seed Determination.} We randomly assign these online players to separate groups, each experiencing a different algorithm to generate an equivalent number of seeds. These algorithms undergo weekly training updates and generate initial predictive scores for player recommendations at the beginning of the week. Nodes with top scores are determined as seeds. The number of seeds is determined by the algorithm with the fewest output predictions.
\par \textit{Seed utilization.} We determine our seeds at the beginning to simulate the IM setting. Throughout the week, these seeds receive priority in recommendation, except when offline, reverting to default recommendations in the absence of entire seeds. This strategy of consistent recommendation is also employed for promotion~\cite{zhang2023capacity,wang2021influence}.
\par \textit{Cool-down mechanism.} We incorporate a ``cool-down'' mechanism to prevent recommendation overload: players who receive ten invitations are excluded from recommendations for 24 hours. This helps prevent player annoyance and evenly distributes social interaction among the seeds, avoiding monopolization of attention by the highest-scoring seeds.


\subsection{Experimental Setting}
\par \textbf{Baseline}. The approaches selected here differ from those in the offline experiments by incorporating a Deep Learning (DL) solution for IM. These SPEX solutions also adhere to our supervised setting for predicting seeds.

\par \textit{IM Competitor with Deep Learning Approach}: We use \textit{GE}~\cite{wang2021influence} as a validated DL baseline due to its proven efficacy.
\par \textit{Graph Solutions}: In our comparative analysis, we utilize \textit{GraphMAE} and \textit{TGN} to explore static and temporal graph analysis, enriching our insights into various graph-based approaches.
\par \textit{Random}: A randomly selected subset of seeds is exposed to default recommendation  generated by a LightGBM model~\cite{ke2017lightgbm} trained to maximize the click-through rate (CTR) for invitation acceptances. This default recommendation serves as a robust business baseline and undergoes thorough online evaluation.

\par \textbf{Metrics}. To evaluate the effect of coverage growth, i.e., network scale growth, we track the daily network scale divided by the accumulation of daily active user (DAU). We use \textit{spread} to denote this network scale growth. In this context, the network edges originate exclusively from the recommendation system, and the DAU accurately reflects the number of available players in a test group. Although detailed DAU figures and any data disclosing DAU are kept confidential, it's important to note that both the number of seeds and the scale of the diffused network are relatively small compared to the DAU. This setup ensures that the regular gaming experience of the majority of players remains undisturbed.

\begin{figure}[t!]
    \centering
    \includegraphics[width=\linewidth]{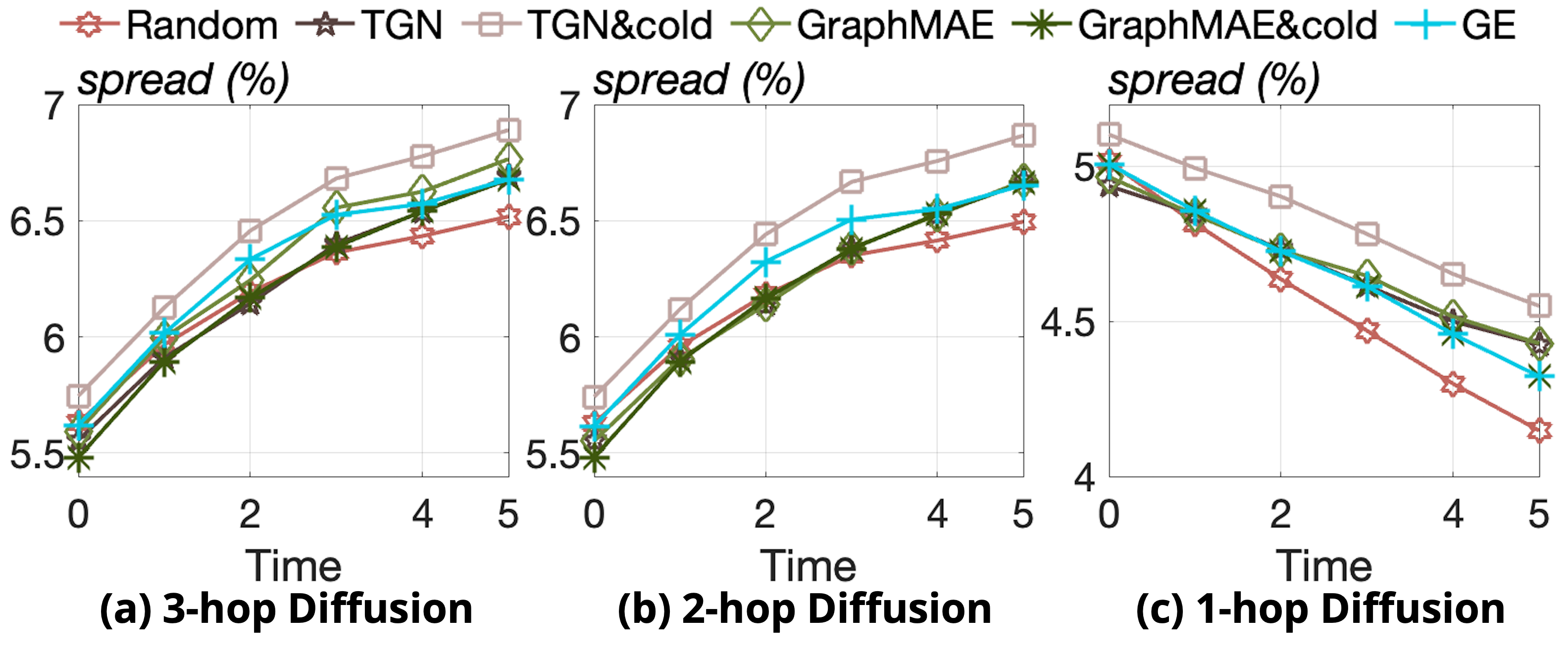}
    \caption{Proportion of network scale over $6$ days.}
    \label{fig:empirical2}
\end{figure}

\subsection{Result Analysis}

\par We begin our results analysis with an overview, followed by targeted investigations into cold-start nodes and an ablation study to validate the cold-start solution. In the illustrations, (x, y) coordinates represent positions from the upper-left corner of each figure.

\par \textbf{Overall Performance}. The diffusion outcomes are detailed in Figure~\ref{fig:empirical2}. Diffusion beyond 3 hops is excluded since the network scale does not exhibit growth beyond this point. Our method, \textit{TGN\&cold}, consistently leads in spreading across both time and hops, achieving a relative improvement of $3.52\%$ in overall statistics. The \textit{random} strategy demonstrates the lowest spread (Time = $4$ and $5$ in Figure~\ref{fig:empirical2}), confirming the observation that social recommendation strategies targeting high adoption rates fall short in expanding network scale. A decrease in $1$-hop diffusion in Figure~\ref{fig:empirical2}(a) indicates that the rate of network diffusion is slower than the accumulation of DAU. This is attributed to the cool-down mechanism, which also serves as evidence that recommending our seeds does not introduce bias into the study of the IM problem.

\begin{figure}[t!]
    \centering
    \includegraphics[width=\linewidth]{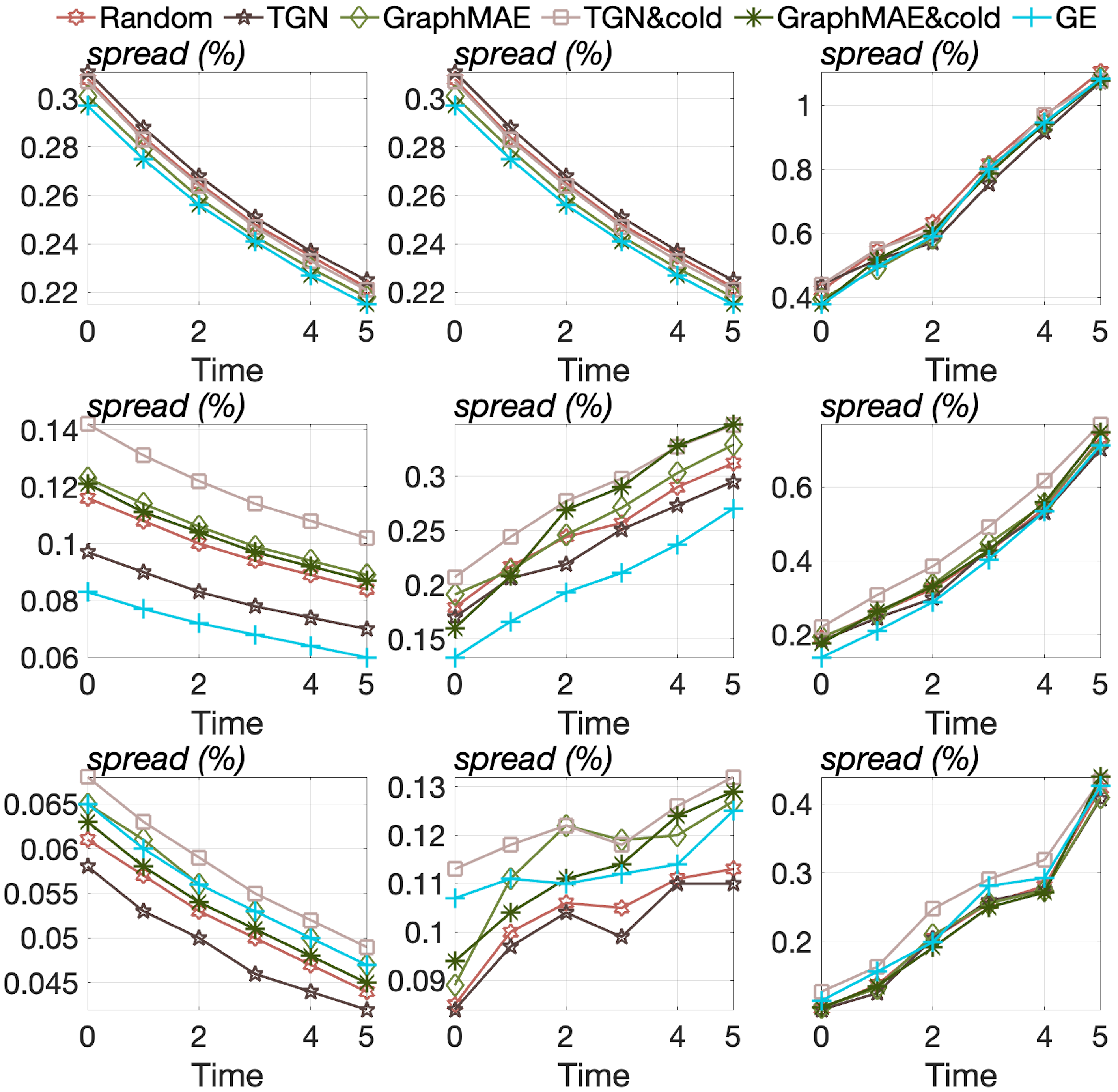}
    \caption{Investigation of degree v.s. hop, where x,y position refers to $\{1,2,3\}\times\{1,2,6\}$. Each figure presents the proportion of network scale over $6$ days.}
    \label{fig:empirical-cold}
\end{figure}

\begin{figure*}[h]
    \centering
    \includegraphics[width=\linewidth]{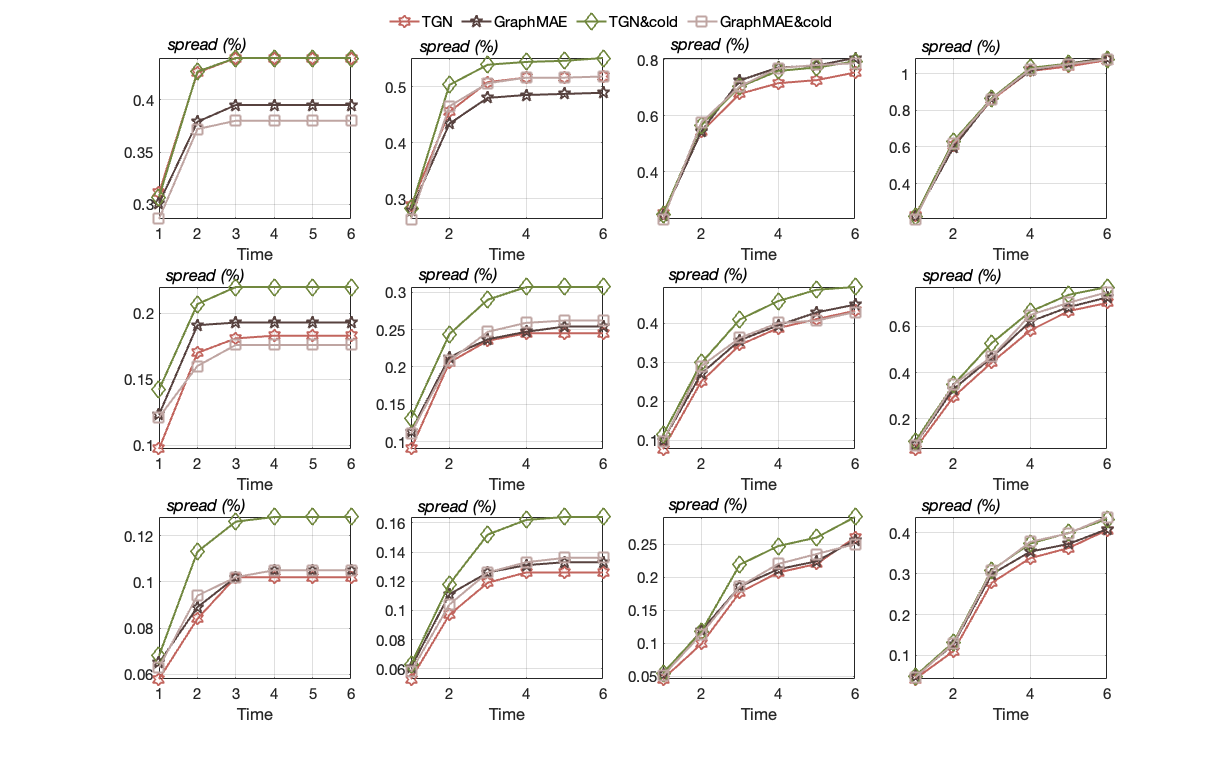}
    \caption{Ablation study of the cold-start solution, where x,y position refers to degrees crossing times, i.e., $\{1,2,3\}\times\{1,2,4,6\}$. Each figure presents the proportion of network scale over $6$-hop diffusion.}
    \label{fig:empirical-ablation}
\end{figure*}

\par \textbf{Cold-Start Treatment}. To address the cold-start issue, we evaluate the enhancements for cold-start nodes, as illustrated in Figure~\ref{fig:empirical-cold}. The analysis focuses on nodes with fewer than three neighbors at the time of seed prediction (i.e., January 28, 2024). \textit{TGN\&cold} either leads or ties in most cases, affirming its effective treatment of cold-start nodes. This is especially evident in cases (2,1), (2,2), (3,1), and (3,2) in Figure~\ref{fig:empirical-cold}, where a clear distinction is observed between the default random strategy and \textit{TGN\&cold}. The overall statistical improvement here is $14.32\%$. To explain, we observe that performance rankings remain largely consistent over time and thus the initial performance determines the overall performance. This observation highlights the significance of the players reached initially, which is impacted by the cold-start issue we particularly focus on to improve the IM problem.

\par Additionally, in Figure~\ref{fig:empirical-cold}, \textit{TGN} exhibits lower performance without the support of our cold-start solution, particularly evident in cases (2,1), (2,2), (3,1), and (3,2). Although \textit{GE} shows comparable performance in Figure~\ref{fig:empirical2}, it also struggles to address cold-start challenges, resulting in the lowest performance in cases (2,1) and (2,2). In contrast to its lower performance in Figure~\ref{fig:empirical2}, the \textit{random} baseline remains comparable throughout Figure~\ref{fig:empirical-cold}. We attribute this to its local optimization strategy, which targets the adoption rate without considering long-term spreading. Consequently, its initially comparable performance does not sustain over time, leading to its lowest performance in Figure~\ref{fig:empirical2}.

\par \textbf{Ablation Study Along Hops}. An ablation study focusing on diffusion across hops is presented in Figure~\ref{fig:empirical-ablation}. Overall, \textit{TGN\&cold} outperforms \textit{TGN} by $15.59\%$, whereas \textit{GraphMAE\&cold} is marginally surpassed by $0.21\%$. This finding is consistent with insights from the offline experiment and clearly demonstrates the benefits of our cold-start solution. Additionally, the increase in spread is particularly evident in the first two hops, aligning with our identification of two-hop IPPs. Beyond the second hop, the spread curves become flat and reach saturation, indicating the potential of our supervised framework to scale up IPP hops. During these two hops, \textit{TGN\&cold} exhibits significantly enhanced propagation power compared with \textit{TGN}, achieving a higher saturation level. In later days, the impact of our cold-start solution diminishes as the propagation accumulates sufficient neighbors, moving past the initial cold-start period. This suggests that the final performance improvement is primarily driven by the initial propagation power provided by our cold-start solution. This is particularly helpful for nodes that reach propagation saturation quickly. However, it fails to break the propagation ceiling for the cold-start nodes in the long term.

\section{Conclusion and Future Work}

\par In this study, we explore supervised learning techniques to tackle the IM problem in temporal graphs, focusing on mitigating the cold-start issue identified through statistical analysis. We first frame the IM problem as a supervised learning task and propose a method for identifying IPPs within multi-relational temporal graphs. To predict these IPPs, we implement a tensorized TGN model, incorporating an innovative cold-start solution. Our approach is validated through offline evaluations across three key metrics, demonstrating both accuracy in IPP prediction and efficiency based on time consumption analysis. Additionally, online A/B testing with a controlled configuration confirms the practical benefits of our approach for fostering network growth. Ablation studies further highlight the efficacy of our solution in addressing the cold-start issue, offering valuable insights and practical contributions to IM in temporal networks.

\par Due to our research scope, we exclude diverse recommendation baselines (e.g., adoption rate~\cite{zhang2023capacity}) and static IM problems~\cite{wang2024fast}. Recent temporal GNN methods~\cite{poursafaei2022towards,luo2022neighborhood} may not scale to our industrial dataset, making TGN a suitable base for advancing temporal networks in scalable IM. Our cold-start solution currently depends on TGN; future work should assess its generalizability across models, explore potential incompatibilities, and enhance adaptability. Nevertheless, future research could address its limited performance in propagation-insufficient domains, such as "who-trusts-whom" networks in Bitcoin trading, and further investigate its behavior in long-hop propagation scenarios.

\section{Ethics Statement}
\par Our empirical study involved conducting online A/B tests. We ensure that our data collection process strictly adhered to privacy and confidentiality standards, and the experimental design is carefully crafted to prevent any negative impacts on players. A potential ethical concern is the abuse of propagation power to influence a large number of players. However, the demonstration of our cool-down mechanism provides a proper guidance to control these unintended propagation effects. Consequently, we believe that our study does not incur any significant ethical issues.

\section{Acknowledgments}
\par This work is supported by grants from the National Natural Science Foundation of China (No. 62372298).

\bibliography{aaai25}

\appendix

\newcommand{\answerYes}[1]{\textcolor{blue}{#1}} 
\newcommand{\answerNo}[1]{\textcolor{teal}{#1}} 
\newcommand{\answerNA}[1]{\textcolor{gray}{#1}} 
\newcommand{\answerTODO}[1]{\textcolor{red}{#1}} 

\section{Paper Checklist}

\begin{enumerate}

\item For most authors...
\begin{enumerate}
    \item  Would answering this research question advance science without violating social contracts, such as violating privacy norms, perpetuating unfair profiling, exacerbating the socio-economic divide, or implying disrespect to societies or cultures?
    \answerYes{Yes, see the Introduction, Conclusion and Future Work.}
  \item Do your main claims in the abstract and introduction accurately reflect the paper's contributions and scope?
    \answerYes{Yes, see the Abstract and Introduction.}
   \item Do you clarify how the proposed methodological approach is appropriate for the claims made? 
    \answerYes{Yes, see the Introduction, Offline Experiment, Empirical Study, Conclusion and Future Work.}
   \item Do you clarify what are possible artifacts in the data used, given population-specific distributions?
    \answerYes{Yes, see the Offline Experiment and Empirical Study.}
  \item Did you describe the limitations of your work?
    \answerYes{Yes, see the Conclusion and Future Work.}
  \item Did you discuss any potential negative societal impacts of your work?
    \answerYes{Yes, see the Introduction.}
      \item Did you discuss any potential misuse of your work?
    \answerYes{Yes, see the Introduction.}
    \item Did you describe steps taken to prevent or mitigate potential negative outcomes of the research, such as data and model documentation, data anonymization, responsible release, access control, and the reproducibility of findings?
    \answerYes{Yes, see the Introduction.}
  \item Have you read the ethics review guidelines and ensured that your paper conforms to them?
    \answerYes{Yes, see the Ethics Statement.}
\end{enumerate}

\item Additionally, if your study involves hypotheses testing...
\begin{enumerate}
  \item Did you clearly state the assumptions underlying all theoretical results?
    \answerNA{NA}
  \item Have you provided justifications for all theoretical results?
    \answerNA{NA}
  \item Did you discuss competing hypotheses or theories that might challenge or complement your theoretical results?
    \answerNA{NA}
  \item Have you considered alternative mechanisms or explanations that might account for the same outcomes observed in your study?
    \answerYes{Yes, see the Offline Experiment and Empirical Study.}
  \item Did you address potential biases or limitations in your theoretical framework?
    \answerYes{Yes, see the Offline Experiment and Empirical Study. Our A/B is under rational and fair setting.}
  \item Have you related your theoretical results to the existing literature in social science?
    \answerYes{Yes, see the Introduction, Motivated Insights and Problem Definition.}
  \item Did you discuss the implications of your theoretical results for policy, practice, or further research in the social science domain?
    \answerYes{Yes, see the Introduction and Conclusion and Future Work.}
\end{enumerate}

\item Additionally, if you are including theoretical proofs...
\begin{enumerate}
  \item Did you state the full set of assumptions of all theoretical results?
    \answerNA{NA}
	\item Did you include complete proofs of all theoretical results?
    \answerNA{NA}
\end{enumerate}

\item Additionally, if you ran machine learning experiments...
\begin{enumerate}
  \item Did you include the code, data, and instructions needed to reproduce the main experimental results (either in the supplemental material or as a URL)?
    \answerNo{No, the empirical study is not reproduceable due to the online environment and the confidential industrial data. However, TGB benchmark can be reproduced, see the open-sourced repository \url{https://github.com/laixinn/TGB}.}
  \item Did you specify all the training details (e.g., data splits, hyperparameters, how they were chosen)?
    \answerYes{Yes, see the Offline Experiment and Empirical Study.}
     \item Did you report error bars (e.g., with respect to the random seed after running experiments multiple times)?
    \answerNo{No, because our experiments take same seed with TGN (2020) to ensure the results are fair and reproduced.}
	\item Did you include the total amount of compute and the type of resources used (e.g., type of GPUs, internal cluster, or cloud provider)?
    \answerYes{Yes, see the Offline Experiment.}
     \item Do you justify how the proposed evaluation is sufficient and appropriate to the claims made? 
    \answerYes{Yes, see the Offline Experiment and Empirical Study.}
     \item Do you discuss what is ``the cost`` of misclassification and fault (in)tolerance?
    \answerYes{Yes, see the Offline Experiment and Empirical Study.}
  
\end{enumerate}

\item Additionally, if you are using existing assets (e.g., code, data, models) or curating/releasing new assets, \textbf{without compromising anonymity}...
\begin{enumerate}
  \item If your work uses existing assets, did you cite the creators?
    \answerYes{Yes, see the Offline Experiment and Empirical Study.}
  \item Did you mention the license of the assets?
    \answerNo{No, because all the tools, algorithms and data we use are publicly available.}
  \item Did you include any new assets in the supplemental material or as a URL?
    \answerYes{Yes, see the URLs in footnotes.}
  \item Did you discuss whether and how consent was obtained from people whose data you're using/curating?
    \answerYes{Yes, see the Offline Experiment and Empirical Study. The data is either publicly available or with consent by players.}
  \item Did you discuss whether the data you are using/curating contains personally identifiable information or offensive content?
    \answerYes{Yes, see the Offline Experiment and Empirical Study. The data is either publicly available or undergoes transformation to keep information confidential.}
\item If you are curating or releasing new datasets, did you discuss how you intend to make your datasets FAIR (see \citet{fair})?
\answerNA{NA}
\item If you are curating or releasing new datasets, did you create a Datasheet for the Dataset (see \citet{gebru2021datasheets})? 
\answerNA{NA}
\end{enumerate}

\item Additionally, if you used crowdsourcing or conducted research with human subjects, \textbf{without compromising anonymity}...
\begin{enumerate}
  \item Did you include the full text of instructions given to participants and screenshots?
    \answerNA{NA}
  \item Did you describe any potential participant risks, with mentions of Institutional Review Board (IRB) approvals?
    \answerNA{NA}
  \item Did you include the estimated hourly wage paid to participants and the total amount spent on participant compensation?
    \answerNA{}
   \item Did you discuss how data is stored, shared, and deidentified?
   \answerNA{NA}
\end{enumerate}

\end{enumerate}

\end{document}